**The Production of Small Primary Craters on Mars and the Moon**

J.-P. Williams*, A. V. Pathare, O. Aharonson

Jean-Pierre Williams*
Dept. Earth and Space Sciences
University of California
Los Angeles, CA 90095, USA
jpierre@mars.ucla.edu
310-825-4414 (office)
310-825-2279 (fax)

Asmin V. Pathare
Planetary Science Institute
Tucson, AZ 85719, USA
pathare@psi.edu

Oded Aharonson
Center for Planetary Science
Weizmann Inst. Of Science
Rehovot, 76100 Israel
oa@caltech.edu

*Corresponding Author

**Abstract**

We model the primary crater production of small ($D$ < 100 m) primary craters on Mars and the Moon using the observed annual flux of terrestrial fireballs. From the size-frequency distribution (SFD) of meteor diameters, with appropriate velocity distributions for Mars and the Moon, we are able to reproduce martian and lunar crater-count chronometry systems (isochrons) in both slope and magnitude. We include an atmospheric model for Mars that accounts for the deceleration, ablation, and fragmentation of meteors. We find that the details of the atmosphere or the fragmentation of the meteors do not strongly influence our results. The downturn in the crater SFD from atmospheric filtering is predicted to occur at $D$ ~ 10-20 cm, well below the downturn observed in the distribution of fresh craters detected by the Mars Global Surveyor (MGS) Mars Orbiter Camera (MOC) or the Mars Reconnaissance Orbiter (MRO) Context Camera (CTX). Crater counts conducted on the ejecta blanket of Zunil crater on Mars and North Ray crater on the Moon yielded crater SFDs with similar slopes and ages (~1 Ma, and ~58 Ma, respectively) to our model, indicating that the average cratering rate has been constant on these bodies over these time periods. Since our Monte Carlo simulations demonstrate that the existing crater chronology systems can be applied to date young surfaces using small craters on the Moon and Mars, we conclude that the signal from secondary craters in the isochrons must be relatively small, as our Monte Carlo model only generates primary craters.

## 1. Introduction

The accumulation of craters on a planetary surface can be used to determine relative ages of areas of geologic interest. Assigning absolute ages is done with modeled impact crater isochrons, a technique that has been developed over several decades (e.g. Hartmann, 1966; Neukum and Wise, 1976; Hartmann, 1999; Neukum and Ivanov, 1994; Hartmann and Neukum, 2001; Hartmann, 2005). For crater ages on Mars, isochrons are derived from the size-frequency distribution (SFD) of craters observed on the lunar maria for which we have dated Apollo samples (Wilhelms, 1987), scaled to account for the ratio of meteoroids at the top the martian atmosphere relative to the Moon's and the differences in gravity and average impact velocity of intersecting orbits. The resulting isochrons yield an expected crater SFD for a given age surface and provide a means of understanding the absolute timescale of major geological and geophysical processes.

The SFD is typically described as a power-law with slope *n*. Deviations from the power-law occur through various processes which, in general, preferentially alter the smaller diameter crater population, making small craters more challenging to use for age-dating surfaces. Because of the frequency at which small craters form however, they provide the ability to discriminate surface ages of geologically young regions and features at a higher spatial resolution where only small craters are available for dating. This level of resolution is required to establish the temporal relation of recent geologic activity on Mars such as gully and landslide formation, volcanic resurfacing, sedimentation, exhumation, dune activity, glaciation and other periglacial landforms, and the possible relation of such features to obliquity cycles of ~$10^7$ y timescale (e.g. Basilevsky et al., 2009; Burr et al., 2002; Hartmann and Berman, 2000; Kadish et al., 2008; Lanagan et al.,

2001; Malin and Edgett, 2000a, 2000b, 2001; Mangold, 2003; Marquez et al., 2004; Quantin et al., 2007; Reiss et al., 2004; Shean et al., 2006; Schon et al., 2009).

The Mars Obiter Camera (MOC) aboard the Mars Global Surveyor (MGS), with ~1.5 m resolution/pixel (Malin et al., 1992), identified 19 fresh craters over a ~6.8 year period (Malin et al., 2006). Data from the High Resolution Imaging Science Experiment (HiRISE) aboard the Mars Reconnaissance Orbiter (MRO) is currently providing image data with up to 25 cm pix$^{-1}$ resolution (McEwen et al., 2007a), and has imaged and confirmed >200 small (< 50 m diameter) fresh impact craters having formed within the last few decades following their discovery by the Context (CTX) camera (Malin et al., 2007) on the same spacecraft (Byrne et al., 2009; Daubar et al., 2010, 2011, 2012; Daubar and McEwen, 2009; Dundas and Byrne, 2010; Ivanov et al., 2008, 2009, 2010; Kennedy and Malin, 2009; McEwen et al., 2007b, 2007c). This offers an opportunity to study the production of small meter-scale craters on the surface of Mars in greater detail and refine isochron models for dating young surfaces on Mars.

In this paper, we model crater populations using a Monte Carlo simulation to explore the primary crater production function at small diameters and the potential influence of present-day atmospheric filtering. We explore ablation, deceleration, and fragmentation of projectiles as they traverse the martian atmosphere and compare our results with the 44 fresh craters reported by Daubar et al. (2013) that have been well constrained by CTX before- and after-images. We then assess the current impact-crater isochron model of Hartmann (2005) and the estimated ratio of meteoroids at the top of the martian atmosphere relative to the top of the terrestrial atmosphere, which is the primary source of uncertainty in isochron models (Hartmann, 2005).

**2. Model**

## 2.1 Meteoroids traversing the atmosphere

Crater populations are modeled assuming a power law distribution of projectiles at the top of the atmosphere and account for possible fragmentation and the dependence of mass and velocity on deceleration and ablation in the atmosphere. The size distribution of projectiles at the top of Mars' atmosphere is adapted from the power-law fit to satellite observations of the annual flux of near-Earth objects colliding with the Earth for objects with diameters < 200 m (Brown et al., 2002): log ($N$) = $c_o$ – $d_o$ log ($D_p$), where $N$ is the cumulative number of bolides colliding with the Earth per year, $D_p$ is the meteoroid diameter in meters, $c_o$ = 1.568, and $d_o$ = 2.70. This SFD has been scaled by a factor 2.6, the nominal ratio of meteoroids at the top of Mars' atmosphere relative to the Moon from the latest isochron model iteration of Hartmann (2005),

The kinetic energy of an object entering the atmosphere is lost to deceleration and ablation. The decelerating force due to aerodynamic drag is (e.g., Baldwin and Sheaffer, 1971; Chyba et al., 1993; Melosh, 1989)

$$\frac{dv}{dt} = \frac{C_D \rho_a A v^2}{2m} + g(z) \sin \theta \tag{1}$$

where $\rho_a$ is the local density of the atmosphere. The parameters $A$, $v$, and $m$, are the cross-sectional area, the velocity, and the mass of the object respectively, $g$ is the local gravitational acceleration at altitude $z$, and $\theta$ is the angle of the trajectory measured from the local horizontal

$$\frac{d\theta}{dt} = \frac{g(z) \cos \theta}{v} \tag{2}$$

Eq. 2 assumes a flat surface geometry and precludes the possibility of projectiles skipping out of the atmosphere. The drag coefficient, $C_D$, is ~1 in the continuum flow regime (Podolak et al., 1988). Heating of the projectile's surface during entry is efficiently shed by ablation

$$\frac{dm}{dt} = \frac{C_H \rho_a A v^3}{2\zeta} \tag{3}$$

where $C_H$ is the heat transfer coefficient and $\zeta$ is the heat of ablation (Bronshten, 1983).

To assess the conditions in which deceleration and ablation become substantial, we employ a heuristic model. Taking the gravity term to be negligible in Eq. (1) for the moment, and the approximation of an exponential density scale height for the atmosphere, $\rho_a = \rho_o \exp(-z/H)$, where $z$ is the altitude, $H$ is the scale height, and $\rho_o$ is the atmospheric density at the surface, the final mass of the meteor, $m_f$, can be related to the initial velocity, $v_i$, and final velocity, $v_f$, by

$$m_f = m_i \exp[-\sigma(v_i^2 - v_f^2)] \tag{4}$$

where $m_i$ is the initial mass of the object at entry and $\sigma = C_H/(2\zeta C_D)$ is the ablation coefficient.

Combining the above equations and solving (e.g., Davis, 1993), the dependence of mass and velocity on deceleration and ablation can be seen (Figure 1). Smaller, faster objects are more readily filtered out by the atmosphere. The condition for significant deceleration can be estimated to occur when the meteoroid mass is equivalent to the column of atmospheric mass it encounters, $m_i \sim \rho_o H A$. Similarly, we can estimate the condition for substantial ablation. Ablation is more efficient at higher velocities as the ablative energy is proportional to the product of the drag force and the traversed distance, or, $dm \cdot \sigma^{-1} \propto \rho_a v^2 A \cdot v dt$. A transition to a high ablation regime will occur when the energy to ablate the entire meteoroid mass, $m \cdot \sigma^{-1}$, is equivalent to the energy required to traverse the atmosphere to the surface at a given velocity, $\rho_o H A \cdot v^2$, neglecting deceleration.

Using these criteria, we can classify the projectiles based on velocity and mass (Figure 2). For large, slow projectiles, $m_f \sim m_i$ and $v_f \sim v_i$, and the projectile will reach the surface relatively unchanged. For smaller, slow objects where $m < \rho_o H A$, and $dm/dt \sim 0$, the final velocity decreases 146   exponentially $v_f \sim v_i \exp\left[-\frac{\rho_o HA}{m \sin\theta}\right]$. Fast meteors, defined as having high initial velocities where

147   $v_i^2 > \sigma^{-1}$, will experience significant ablation, $m_f < e^{-1} \cdot m_i$. Large, fast meteors will survive

148   complete ablation if $m\sigma^{-1} < \rho_o HA \cdot v^2$, where the ablation will be limited by deceleration. These

149   intermediate projectiles define the wedge shaped region in Figure 2a. Our taxonomic boundaries

150   based on initial projectile masses and velocities are derived from our numeric modeling for a

151   range of projectile masses and entry velocities (Figure 2).

**2.2 Monte Carlo simulation**

153   We generated model crater populations using a Monte Carlo simulation employing the power-

154   law distribution of meteoroids entering the top of the atmosphere from the previous section and

155   the normalized distribution of entry velocities at Mars (Bland and Smith, 2000; Davis, 1993;

156   Flynn and McKay, 1990; Popova et al., 2003):

$$F(v_i) = 0.0231 v_i \exp\left[-\left(\frac{v_i - 1.806}{8.874}\right)^2\right] \qquad (4)$$

157   where $v_i \geq 5$ km s$^{-1}$ (the escape velocity of Mars) with a mean velocity of 10.2 km s$^{-1}$. Following

158   Love and Brownlee (1991), the probability distribution of entry angles is taken to be $\sin(2\theta)$,

159   which has a maximum at 45° and drops to zero at 0° and 90°. A distribution of 5 groups of

160   material types derived from the relative observed number of objects entering the terrestrial

161   atmosphere is taken from Ceplecha et al. (1998). The material groups differ in their ability to

162   penetrate the atmosphere, with the average ablation coefficient, $\sigma$, and bulk density, $\rho_m$, for each

163   group listed in Table 1. Model results for ordinary chondrites are shown in Figure 3 for a single

164   initial entry angle of 45°, demonstrating a good match with the results shown in Figure 1.

165       The resulting crater volumes from the projectiles impacting the surface can be described

166   by a scaling law that relates impact velocity and projectile and target characteristics (see Melosh

(1989) for a review). In converting the crater SFD derived from the lunar surface to Mars, Hartmann (2005) assumes crater diameters, $D$, scale with impact energy as $E^{0.43}$ and with gravity as $g^{-0.17}$ to account for differences in mean impact velocity and gravitational acceleration between the two bodies. For small projectiles, the depth of excavation by an impact is small enough that lithostatic stresses are small relative to the characteristic yield stress of the regolith. In such cases the resulting transient crater diameter is determined by the yield strength, $\bar{Y}$, of the target material ("strength scaling"), and no longer scales with gravity. The transition from strength scaling to gravity scaling occurs when $\bar{Y} \sim \rho_t g R_p$ where $R_p$, the projectile radius, is taken to be the characteristic depth and $\rho_t$ is the target density (Holsapple, 1993). For a regolith of density 2000 kg m$^{-3}$ and yield strength of 0.1 – 1 MPa, the transition between strength and gravity dominated regimes occurs at $D \sim 27 - 270$ m. This implies that the 1 – 10 m scale fresh craters discovered by spacecraft over the last decade (Malin et al, 2006; Daubar et al., 2013) are predominately in the strength scaling regime. Strength scaling results in a reduction in crater volume relative to gravity scaled craters. As a consequence, the distribution of crater diameters would be expected to be shallower than that predicted by the isochrons of Hartmann (2005) at the smallest sizes, which assume all crater diameters scale with gravity.

The cratering efficiency, $\pi_v = \rho_t V/m_f$, where $\rho_t$ is the target density and $V$ is the transient crater volume, is proportional to (Holsapple, 1993):

$$\left(\pi_2 + \pi_3^{\frac{2+\mu}{2}}\right)^{-\frac{3\mu}{2+\mu}} \qquad (4)$$

assuming impactor and target densities are the same, where $\mu$ is an empirical constant (ranging from 1/3 – 2/3). The parameters $\pi_2$ and $\pi_3$ are dimensionless numbers describing the cratering efficiency for gravity scaling and strength scaling, respectively, where $\pi_2 = gR_p/v_f^2$, the ratio of the lithostatic pressure at a depth equivalent to the projectile radius and the initial dynamic 189   pressure generated by the impact, and $\pi_3 = \bar{Y}/\rho_t v_f^2$, the ratio of effective target yield strength to

190   the initial dynamic pressure (Holsapple, 1993). For smaller projectiles, the $\pi_2$ term becomes

191   negligible and $\pi_v$ becomes constant with impactor size, depending only on velocity and the

192   material strength of the target. At large impactor sizes, $\pi_2$ dominates and the cratering efficiency

193   is then dependent not only on velocity, but impactor size. The expression for transient crater

194   volume is (Holsapple, 1993; Richardson et al., 2007; Dundas et al., 2010):

$$V = K_1 \left(\frac{m_f}{\rho_t}\right) \left(\pi_2 \left(\frac{\rho_m}{\rho_t}\right)^{\frac{1}{3}} + \pi_3^{\frac{2+\mu}{2}}\right)^{-\frac{3\mu}{2+\mu}} \qquad (5)$$

195   where $K_1$, like $\mu$ and $\bar{Y}$, are experimentally derived properties of the target material. Taking the

196   transient crater volume to be $(1/24)\pi D_t^3$, assuming the transient crater depth is roughly 1/3 its

197   diameter, $D_t$ (Melosh 1989; Schmidt and Housen, 1987), this equation can be expressed in terms

198   of final crater diameter

$$D = 1.3 K' D \left(\frac{\rho_m}{\rho_t}\right)^{\frac{1}{3}} \left(\pi_2 \left(\frac{\rho_m}{\rho_t}\right)^{\frac{1}{3}} + \pi_3^{\frac{2+\mu}{2}}\right)^{-\frac{\mu}{2+\mu}} \qquad (6)$$

199   where $K' = (4K_1)^{1/3}$ for a spherical meteoroid and the factor 1.3 is the ratio of the final rim-to-rim

200   diameter and the transient crater diameter (Holsapple, 1993). We adopt values for Mars

201   consistent with dry desert alluvium of $\bar{Y} = 65$ kPa, $\mu = 0.41$, $K_1 = 0.24$, and $\rho_t = 2000$ kg m$^{-3}$

202   (Holsapple, 1993; Holsapple and Housen, 2007)

203         A minimum impact velocity threshold of 0.5 km s$^{-1}$ is selected for the formation of

204   explosive impact crater formation as projectiles with lower speeds are unlikely to generate a

205   shock wave of sufficient magnitude to crush the target material and excavate an explosive crater

206   (Popova et al. 2003). The definition of this velocity is not sharply defined and will depend on

207   target material properties. Pressure wave velocities in competent basalt are typically ~4.5 – 6.5

208    km s$^{-1}$, with values an order-of-magnitude lower for loose, unconsolidated soils. Apollo seismic

209    investigations found velocities to be ~0.1 – 0.3 km s$^{-1}$ in the upper 100 m of the lunar regolith

210    (Kovach and Watkins, 1973; Watkins and Kovach, 1973).

211        The initial altitude of the modeled projectiles is 100 km. The gas density at this altitude is

212    reduced by four orders-of-magnitude relative to the atmospheric density at the planet's surface,

213    and thus deceleration (Eq. 1) is reduced by four orders-of-magnitude relative to when the

214    meteoroid is near the surface and is therefore initially negligible. This altitude also represents the

215    approximate transition between the free molecular flow regime and the continuum regime (i.e.

216    Knudsen number ~ 0.1, depending on the object size, for a mean free path of ~ 1 cm).

217    **2.3 Aerodynamic breakup**

218     The fragmentation of meteoroids during their flight through the atmosphere likely influence the

219    resulting SFD of smaller diameter craters, as more than half (56%) of the >200 fresh craters

220    observed by HiRISE  are comprised of clusters of individual craters (Daubar et al., 2013).

221    Numerous additional examples of crater clusters have been observed on the surface of Mars

222    (Popova et al., 2007) implying that the fragmentation of meteoroids that penetrate deep into the

223    atmosphere is a common phenomenon on Mars. The aerodynamic breakup of a projectile is

224    expected when the dynamic pressure experienced during entry, $\rho_a v^2$, exceeds the bulk strength,

225    $\sigma_m$, of the projectile. Bolides entering the terrestrial atmosphere display a wide range of strengths

226    inferred from the broad range of observed breakup event altitudes with bulk strengths shown to

227    be ~0.1 – 10 MPa (Ceplecha et al., 1998; Popova et al., 2011). The estimated bulk strengths are

228    low compared to the tensile strength of recovered samples (typically 1-10 %), implying fractures

229    and other zones of weakness within the bodies determine bulk strength rather than material

230    composition (Popova et al., 2011). Dynamic pressures experienced by projectiles entering the Martian atmosphere will commonly exceed 1 MPa and therefore fragmentation should be expected, albeit at a lower altitude than observed in the terrestrial atmosphere.

Larger, faster objects experience greater dynamic pressure during flight (Figure 4). Though there is no explicit size dependence on dynamic pressure, smaller objects begin to ablate and decelerate higher in the atmosphere, and therefore experience smaller peak loading at higher altitudes than equivalent larger objects. This implies a size dependence on fragmentation from entry dynamics alone. Further, it is anticipated that larger objects are inherently weaker as they likely contain a greater number of defects, and a power law relation based on statistical strength theory (Weibull, 1951) between bulk strength and sample mass is typically assumed in models (e.g. Artmieva and Shuvalov, 2001; Popova et al., 2003; Svetsov et al., 1995). However, recent analysis of meteorite fall observations in the terrestrial atmosphere suggests that, at least for stony objects, the relation between bulk strength and sample mass may be weaker (Popova et al., 2011). A possible size dependence on fragmentation was reported by Ivanov et al. (2010) based on a preliminary analysis of ~70 fresh craters, and is also exhibited by the 44 fresh craters observed by Daubar et al. (2013), 25 of which consist of two or more craters (57%). This percentage increases for craters with $D > 5$ m to 73% and decreases to 41% for craters with $D < 5$ m.

**3. Results**

**3.1 Fragmentation**

The effects of fragmentation on the crater SFD is explored with the model. We include aerodynamic break up of meteoroids in the simulation allowing fragmentation to occur when the dynamic pressure exceeds a threshold bulk strength, $\sigma_m = \rho_a v^2$ (Figure 5). Fragmentation is assumed to occur as a single event; however, multiple fragmentation events at different points in a meteoroid's trajectory are sometimes observed during the flight of terrestrial fireballs. Catastrophic fragmentation (a point source type explosion with thousands of fragments) also occurs occasionally. We constrain the number of fragments to 2-100 and each fragment is followed individually to the termination of its flight in order to assess how fragmentation alters the resulting SFD. The number of fragments generated is assumed to follow a power-law probability: we adopt a power-law slope of -1.5 for our nominal model and adjust this value to explore the sensitivity of our results on this exponent. Decreasing the power-law slope results in a more uniform distribution of fragment numbers, while increasing the slope produces a distribution favoring fewer, larger, fragments. A cascade of fragment sizes is then generated by iteratively subtracting random mass fractions from the initial mass which typically results in a small number of larger fragments and many smaller fragments. We additionally explore selecting mass fractions with a uniform random distribution of fragment sizes.

The model is run with $10^6$ initial projectiles with a minimum allowable projectile diameter of 2 cm. A significant number of these events do not survive, but this minimum value was found to resolve the atmospheric downturn in the resulting crater SFD allowing the smallest permissible craters to form. The same bulk strength is selected for all the projectiles, and adjusted until the ratio of crater clusters to total impacts for $D > 5$ m is similar to the 73% observed for the 44 craters reported by Duabar et al. (2013). We find that a bulk of strength of $\sigma_m$ = 0.65 MPa best reproduces the observations, as shown in Figure 6, which plots histograms in $\sqrt{2}$ bins of the resulting crater diameters. Both the observed and model craters are dominated by crater clusters at $D > 5$ m of approximately the same fraction. Effective diameters are estimated for the clusters by $D_{eff} = (\Sigma_i D_i^3)^{1/3}$, which represents the diameter of an equivalent crater due to a non-fragmented meteoroid (Malin et al., 2006; Ivanov et al., 2008).

There is a downturn in the SFD of the craters observed by Daubar et al. (2013) at $D \sim 4$-5 m. However, the model craters continue to increase in number at smaller diameter bins down to the $D = 0.24$ m bin in Figure 6. The majority of craters have small diameters < 1m, and the fraction of clusters in each bin becomes negligible, with the total fraction of all clusters comprising only 4% of the total model crater population. This implies that the atmospheric downturn has not been observed in the fresh craters, and that a significant number of craters with $D < 4$-5 m have not been identified. The discovery of fresh craters relies on the detection of dark spots in CTX images in dusty regions. A swarm of fragments will disturb a larger surface area creating, in general, larger dark spots. The predominance of single craters at $D < 5$ m indicates that detection will be more challenging as dark spots become inherently smaller relative to $D$ at these sizes, partly explaining the downturn in the observed SFD at 4-5 m.

One might conclude that fragmentation, being only 4% of the events, may not be an important process in shaping the SFD. However, crater clusters begin to dominate the SFD at the very diameters that can be reliably detected by CTX, and therefore the effects on absolute ages derived from crater production functions are considered. Comparing the SFD of the model craters with craters produced by the same projectiles with fragmentation suppressed (Figure 7), a difference in age of 7% is obtained when fitting the Hartmann (2005) production function for $D > 4$ m. The influence of fragmentation on the crater SFD can be increased by taking the probability of the number of fragments generated when $\sigma_m$ is exceeded as random and uniform, instead of following the -1.5 slope power law (Figure 5b), and a uniform distribution of fragment masses. In this case we obtain a factor ~0.69 difference in age relative to the age derived if no fragmentation is allowed to occur (Figure 7b). This represents the case we have considered to have the largest likely influence on the SFD.

The change in the crater SFD with fragmentation can be attributed to the mass dependence on ablation and deceleration and corresponding changes in crater scaling, which is not accounted for when calculating effective crater diameters, $D_{eff}$. Individual fragments experience greater deceleration and ablation per unit volume than the larger parent meteoroid. If enough of the total mass is shifted into a high deceleration/ablation regime when the parent object is broken into individual fragments, $D_{eff}$ may no longer be representative of the crater generated by an equivalent unfragmented meteoroid. If the ratio of $D_{eff}$ to the actual $D$ resulting from the unfragmented projectile is $<< 1$, then $D_{eff}$ is a poor representation of the equivalent unfragmented crater. In general, $D_{eff}/D < 1$ when $D_{eff} \lesssim 10$ m for the nominal fragmentation case, with the potential reliability of $D_{eff}$ decreasing with size (Figure 8). Scatter in $D_{eff}/D$ values is largely due to variations in meteoroid velocities with lower $D_{eff}/D$ for faster objects. Some events result in $D_{eff}/D$ values slightly greater than 1, predominately at $D_{eff} \gtrsim 10$ m, due to the dependence of impactor size on gravity scaling where crater efficiency decreases with increasing impactor size. Consequently, the crater efficiency can be larger for the fragments than for the parent projectile is some cases.

**3.2 Impact-crater isochrons**

The technique of dating the Martian surface using predicted crater SFDs for well-preserved surfaces using impact-crater isochrons has been developed over several decades going back to the early lunar exploration of the 1960s (e.g. Hartmann, 1966; Neukum and Wise, 1976; Hartmann, 1999; Neukum and Ivanov, 1994; Hartmann and Neukum, 2001; Hartmann, 2005). Radiometric and exposure ages from Apollo and Luna samples, correlated with crater populations, have anchored the lunar cratering chronology. Scaling lunar isochrons for Mars to account for the ratio of meteoroids at the top of the martian atmosphere relative to the Moon and the difference in gravity and average impact velocity provides a means of understanding the absolute timescale of major geological and geophysical processes on Mars with order of magnitude accuracy. Degradation of craters over time will result in a departure in the crater population from the isochrons. An inflection at $D\sim1.5$ km separates the isochrons into two branches with different power-law exponents. This inflection, initially observed in the lunar craters, divides the crater distribution into a shallow branch with a slope of ~2 and a steep branch of ~3 on a log-differential plot. In his latest iteration of Mars isochrons, Hartmann (2005) included an atmospheric model (Popova et al., 2003) to encapsulate the effects of atmospheric entry of meteors resulting in curvature in the isochrons at sub-km crater diameters.

During impact, ejecta traveling large distances from a crater are capable of generating isolated secondary craters that could potentially be mistaken for primary craters. What fraction of observed crater populations are primary versus secondary, and the consequence of unidentified secondary craters in dating surfaces, is a matter of debate. McEwen et al. (2005), and McEwen and Bierhaus (2006), argue that small crater populations are dominated by secondary craters, rendering smaller craters unreliable for dating. Hartmann (2005, 2007) however points out that in formulating isochrons, no attempt is made to exclude all secondaries and contends that the accumulation of secondary craters are not "contamination" but rather part of the signal. A population of primary plus secondary craters will have a steeper SFD power law slope than a population devoid of secondary craters, and McEwen et al. (2005) has proposed that the increase in slope of the steep branch of crater SFDs is the result of secondaries.

Our approach of using the observed projectile production function at the top of the terrestrial atmosphere provides an opportunity to independently test the veracity of lunar and martian isochrons. We can also constrain the potential influence of secondary craters, as our Monte Carlo model only produces primary crater populations. If secondary craters are a substantial fraction of crater populations, then our model crater SFD should deviate in both slope and age from the isochrons. Additionally, since we use the present-day flux of projectiles, any significant deviation in age estimates can be attributed to a variation in impact rate over time.

We test our model against crater counts conducted on the proximal ejecta of the crater Zunil. The crater Zunil is selected because it is likely the last $D = 10$ km scale crater to form on Mars (McEwen et al., 2005; Hartmann et al., 2010) with an estimated age of ~1 Ma based on the isochrons of Hartmann (2005), so the distribution of small craters on its ejecta are likely to be predominately primary. Crater counts were conducted on a ~5 km$^2$ area north of the crater rim (Figure 9). The resulting crater SFD is consistent with an age ~1 Ma, similar to counts conducted by Hartmann et al. (2010). We then modeled a population of craters over a 1 Myr period of time for the same surface area as our crater counts. The resulting SFD from both the counts and the Monte Carlo model overlap for bins $D > 2$ m and fall near the 1 Myr isochron of Hartmann (2005) with a similar slope on a log-differential plot (Figure 9b). Cumulative crater frequencies are also shown (Figure 9c), and yield effective crater retention ages of $0.903 \pm 0.077$ Ma and $0.820 \pm 0.071$ Ma for the counts and the model respectively. Effective diameters from the crater counts were determined by assuming any craters within 40 m of each other were part of a cluster. For comparison, the SFD of the individual observed craters and model craters result in a modest upward revision of ages to $1.03^{+0.078}_{-0.077}$ Ma and $0.966 \pm 0.074$ Ma respectively (Figure 9d). As crater densities increase on a surface, the identification of individual clusters becomes more difficult, making age determinations more challenging for smaller craters on older surfaces.

We also conducted a similar Monte Carlo comparison with the 44 fresh CTX-CTX detected craters. A crater population is modeled for a corresponding area and time represented by the CTX craters. The model results in a SFD overlapping with the 1 yr isochron of Hartmann (2005). However, the fresh craters fall below the isochron as noted by Daubar et al. (2013). The largest crater, $D = 33.8$ m, is consistent with a single $D \sim 30$ m crater predicted to form annually for the scaled observation area. However, smaller diameter bins are inexplicably under-populated (Figures 9,10) Note that the atmospheric downturn is expected to occur at much smaller diameters (sub-meter), and fragmentation will not substantially deplete the SFD at these smaller sizes (Figure 6b). A lack of a contribution from secondaries to the isochrons also cannot explain the discrepancy, as our model predicts a primary SFD with a similar slope and age as the isochron.

Fresh craters discovered using MOC by Malin et al. (2006) also have an apparent deficit of craters at all bin diameters when scaled to the 1-yr isochron of Hartmann (2005). The cumulative distribution of fresh craters, including both CTX-CTX and MOC detections, is shown in Figure 10. The MOC SFD has been scaled to an equivalent time/area for comparison. The MOC camera, with a lower resolution, did not detect any craters below $D \sim 10$ m and detected fewer craters in all bin sizes that overlap with the CTX craters. All bin sizes (except for the largest CTX crater) fall below the isochron, and this discrepancy is greater in the lower resolution dataset. This indicates that observational bias may be the best explanation for the difference between the predicted SFD and the observed SFDs. If this is the case, then over time, as additional, larger diameter craters are identified by CTX/HiRISE, the SFD will start to overlap with the isochron at larger diameter bins since the larger craters are more likely to create detectable dark spots on the surface. This would then suggest that fresh craters were undercounted by MOC at all diameters observed.

**3.3 Lunar crater counts**

As an additional test, we compare results from the model with crater counts of small craters on the moon (Figure 11). The ejecta of North Ray crater was selected for its relatively young age, which has been constrained by cosmic ray exposure ages of Apollo 16 samples to be ~50 Ma (Arvidson et al., 1975). Our small (0.1 km$^2$) study area only contains craters $D \leq 22$ m. Applying the cratering chronology of Neukum et al. (2001), an age of $58.9 \pm 11$ Ma is derived from the resulting cumulative SFD, consistent with the age derived by Hiesinger et al., (2012) using a larger overlapping area.

A population of projectiles is then generated, again using the size distribution of small near-Earth objects colliding with the terrestrial atmosphere (Brown et al., 2002) and the velocity distribution of Marchi et al. (2009), for a period of 58.9 Ma. This results in a total of 42895 craters for the 0.1 km$^2$ surface area using projectiles >0.02 m diameter. Crater scaling parameters appropriate for the lunar regolith are employed assuming a regolith strength of 10 kPa, $K_1 =$ 0.132, and $\rho_t = 1500$ kg m$^{-3}$ (Holsapple, 1983; Holsapple and Housen, 2007; Vasavada et al., 2012). The resulting SFD of the Monte Carol model craters results in a similar age of $57.2 \pm 11$ Ma (Figure 11). This demonstrates the usefulness of our approach, as we are able to reproduce a ~50 – 60 Ma surface age on the Moon using a limited area containing only small craters. If craters in this area were predominately secondary craters at these diameters, then our Monte Carlo model should have predicted a much younger surface age. A similar approach was taken by Ivanov (2006) to demonstrate that small craters on the Moon must be predominately primaries on young surfaces ($\leq 100$ Ma), consistent with our model.

Our ability to reproduce isochron ages using a small area with a limited range of diameters not only validates the applicability of small craters for dating young < 60 Ma surfaces, but also indicates that the cratering rate has on average been constant over this period (Ivanov, 2006). How secondary craters alter the SFD and whether such small craters can be reliably used for age dating surfaces in all cases is beyond the scope of this work, but our Monte Carlo approach does independently confirm that crater-count chronometry systems can accurately date the youngest surfaces on both the Moon and Mars.

**3.4 Sensitivity analysis**

The sensitivity of our results to changes in model parameters is evaluated (Table 2). Our nominal case results in a good fit to the Hartmann (2005) annual isochron for $D \geq 20$ m where the model produces a SFD that results in an age of $1.03 \pm 0.017$ yr (Figure 12). The Monte Carlo model does not produce as close a match at smaller sizes where the SFD is best fit with an age of $0.734 \pm 0.0018$ yr for $D = 4 - 20$ m. This may result from a greater atmospheric influence on the meteoroids at smaller sizes than predicted by the atmospheric model of Popova et al. (2003) that is incorporated into the isochron, or could be due to Hartmann (2005) assuming gravity scaling only which would also result in less downturn in the isochron as crater scaling transitions into the strength regime (Williams et al. 2010).

The mass of the atmospheric column that a meteoroid must traverse will depend on the location of the event as the elevation of the surface can vary substantially. The highest elevation, the summit of Olympus Mons, is ~21 km above datum, while at the other extreme, the interior of Hellas Basin, is ~8 km below. This represents a range of over 2.5 atmospheric scale heights. One of the most striking characteristics of the topography is that its distribution is bimodal due to the north-south dichotomy in crust thickness (Smith et al., 1999). The more heavily cratered highlands in the southern hemisphere have an average elevation of ~1.5 km while the distribution of lowland elevations narrowly peak ~5.5 km lower than the highlands (Aharonson et al., 2001). This translates to an increase in average atmospheric density at the surface by a 438 factor 1.6 between the southern highlands, $\rho_o$ = 0.0135 kg m$^{-3}$, and the northern lowlands, $\rho_o$ =

439 0.0222 kg m$^{-3}$. Taking the same model projectiles and generating craters for the two atmosphere

440 densities representative of the average elevations of the highlands and lowlands, two crater SFDs

441 are generated for comparison (Figure 12b). Craters $D \geq 20$ m result in ages 1.14 yr and 0.993 yr

442 respectively, bracketing the nominal model results. This is a change of +10.7% and -3.6%.

443 Fitting the smaller diameters, $D = 4 - 20$ m, results in ages 0.92 yr and 0.69 yr, a change of

444 25.3% and -6.0%. The greater sensitivity at the small crater diameters is to be expected as

445 smaller projectiles are more sensitive to atmospheric conditions.

446 Increasing and decreasing the ablation coefficients, $\sigma$, has a similar effect as changing the

447 atmospheric surface density, as an increase in the rate of mass loss results in smaller objects

448 which in turn results in greater deceleration. Models for the same projectiles with ablation

449 coefficients half the nominal values and double the nominal values were run (Figure 12c). The

450 derived ages for craters $D \geq 20$ m increased by 13.5% and decreased by 18.2% for $\sigma \times \frac{1}{2}$ and

451 $\sigma \times 2$ respectively, and similarly for craters $D = 4 - 20$ m increased by 28.9% and decreased by

452 30.4%.

453 The model assumes that the different taxonomic types of fireballs distinguished from the

454 larger Earth-entering meteoroids (Ceplecha et al., 1998; Popova et al., 2003) have the same

455 fractional distribution at Mars. These types of fireballs differ in their ability to penetrate the

456 atmosphere and therefore if the proposed fractions of projectiles are substantially different on

457 average from that observed in the terrestrial atmosphere (Ceplecha et al., 1998), the results will

458 change. For example, a greater percentage of higher density objects will experience less

459 deceleration and result in larger craters for a given population of projectiles. The smallest craters

460 in the simulation, cm-scale, are formed exclusively by iron objects due to their high density

461 (7800 kg m$^{-3}$). Iron meteors however are a small fraction of the total population (3%). The

462 majority of the objects are Ordinary Chondrites (29%), Carbonaceous Chondrites (33%), and

463 Cometary Material (26%). None of the Soft Cometary Material objects, with a density of 270 kg

464 m$^{-3}$, survive passage through the martian atmosphere. To demonstrate the sensitivity of the

465 model results to different projectile compositions, we run the simulation assuming pure

466 compositions of the three predominant types: Ordinary Chondrites, Carbonaceous Chondrites,

467 and Cometary Material (Figure 12d). The Ordinary Chondrites, with a density greater than the

468 average combined density, results in a factor >2 increase in age, while the Carbonaceous

469 Chondrites reduce the age by 37.3% for $D = 4 - 20$ m and 15.3% for $D \geq 20$ m. A meteor

470 population comprised entirely of Cometary Material results in substantially younger age

471 estimates, 0.0466 and 0.15 yr for the small and large diameters respectively, younger than the

472 ages derived from either the MOC or CTX-CTX craters, with a slope that deviates from the

473 isochron increasingly at smaller $D$. The crater counts from Zunil in Figure 9c are shown in

474 Figure 12d for comparison, scaled to the 1yr isochron, showing a slope consistent with a

475 population of objects with average density similar to the nominal values assumed and

476 inconsistent with a substantial fraction of icy, low-density impactors.

477 The crater scaling becomes increasingly sensitive to the target material properties as

478 projectile diameters decrease with weaker material producing a larger crater for a given

479 projectile. Details of the target properties, particularly the effective strength, may not be well

480 constrained and can vary by location, feature, or geologic unit. Dundas et al. (2010) has

481 demonstrated how this can complicate geologic interpretation as surfaces of the same age, but

482 differing material properties, may yield different ages. We explore the sensitivity of the model to

483 a range of target properties.  Figure 12e shows the results for the same population of projectiles for four different targets. We take the lunar regolith of section 3.3 as a lower bound on target strength representing a low density, dry, moderately porous soil, and soft rock and hard rock from Holsapple (1993) for two targets with greater strength and density than our nominal Mars regolith. For the rock materials, $\mu = 0.55$, the higher value reflecting lower porosity and therefore less energy dissipation (Holsapple and Housen, 2007), and $K_1 = 0.20$. The densities are assumed to be 2250 and 2500 kg m$^{-3}$ and the material strengths 7.6 and 18.0 MPa for the soft rock and hard rock respectively (Holsapple, 1993; Richardson et al., 2007). The hard rock gives an upper bound representative of competent, young basalt that lacks a substantial regolith cover.

      The discrepancy in ages between resulting crater SFDs of the different target materials increases with decreasing $D$. The populations begin to converge $D \gtrsim 100$ m. For $D > 20$ m, the lunar regolith model yields a similar age to the nominal model however for smaller craters, the weaker target material results in larger craters and a corresponding increase in estimated age, 0.903 versus 0.734 yr. The stronger rock materials yield smaller craters and the decrease in estimated ages with the rock materials results in SFDs that overlap with the observed CTX-CTX and MOC fresh craters (Figure 12e). For our model to be consistent with the SFD of the fresh craters, the majority of the craters would have to have formed in relatively strong, competent rock. Many of the craters are on Amazonian units in the Tharsis region which is dominated by Amazonian lava flows, however, the largest CTX crater, $D = 33.8$ m, formed on the southeast flank of Pavonis Mons on a relatively young Amazonian volcanic unit (Scott et al., 1998). This crater is consistent with Hartmann 2005 which predicts an annual $D \sim 30$ m diameter crater and its presence in the fresh crater population cannot be attributed to the crater having formed in weaker target material relative to the other fresh craters. It is tempting to dismiss the single crater as a statistical outlier. However, it is incorrect to interpret the $\sqrt{n}$ error bars in comparison to

model isochrons as these error bars indicate the 1-$\sigma$ range of likely *observations* not *models* given the actual number seen. If we wait an additional year, we may observe no additional events of this size. However the real flux of events of this size clearly cannot be 0 given that a single event has been observed and so models with exceedingly low formation rates must be formally rejected (Aharonson, 2007).

If continued observations of fresh CTX-CTX craters verify that the current observed SFD is indeed an accurate reflection of the present-day impact crater population, then our model demonstrates that it is possible to constrain the fraction of impactor material types and average target material properties for Mars. For example, a population of impactors with average densities similar or higher than that observed for terrestrial fireballs could be ruled out as they are not consistent with the current CTX-CTX observations.

**4. Discussion**

The fact that the slope of the Hartmann (2005) isochrons are the same as that derived in our study using the present-day annual flux of terrestrial fireballs implies that the isochron system does not incorporate many unrecognized secondary craters. This is quite surprising considering McEwen et al. (2005) identify ~$10^7$ secondary craters generated by the Zunil crater forming impact event at distances up to 3500 km away from the primary crater. This implies that small secondaries should be abundant on the Martian surface. We specifically selected young surfaces to conduct counts that likely reflect a primary crater population so it is not surprising that our counts have similar slopes to our model. The isochron system however, is derived from counts conducted on lunar maria with average ages of 3.5 Ga, surfaces that have had ample time to accumulate secondaries. Obvious secondaries were not included in those counts and presumably only unrecognizable distant secondaries were included. It remains unclear how secondaries are influencing the crater statistics as we do not identify a secondary signal. We have not attempted a systematic survey to identify a range of crater diameters and surface ages where secondaries are influencing the crater SFD, which is beyond the scope of this study, and caution in using small craters remains warranted as this remains an unresolved issue. However, we do demonstrate that for young surfaces that are not likely contaminated by secondaries, useful ages can be derived from crater counts using small craters.

**5. Conclusions**

Taking the size distribution of the observed annual flux of terrestrial fireballs, scaling the number by 2.6 for Mars, the nominal value of Hartmann (2005), and appropriate velocity distributions for Mars and the Moon, we are able to generate crater populations with a remarkably similar SFD to that predicted by the isochrons of Hartmann (2005) for Mars and Neukum et al. (2010) for the Moon. Our Monte Carlo model generates a population of primary craters, implying that the isochrons do not contain a significant secondary crater signal as large numbers of secondaries would result in a steeper isochron slope than predicted by our model. This further implies that the inflection from the shallow branch to the steep branch is not attributable to the addition of unidentified secondary craters. Crater counts conducted on the ejecta blanket of Zunil and North Ray crater yield SFDs with similar slopes and ages as our model— ~1 Ma and ~58 Ma, respectively, for Mars and the Moon—demonstrating that the cratering rate on average has been fairly constant on these bodies over these time periods.

      We find that the details of the atmospheric model for Mars do not substantially alter the results. The Hartmann (2005) isochrons include an atmospheric model (Popova et al., 2003), and differences between the lowlands and highlands do not result in significant deviations from the Hartmann isochron, nor do variations in assumed ablation coefficients. Fragmentation did not

significantly alter the model results and smaller projectiles ($D < 5$ m) typically were decelerated enough to avoid fragmentation. The atmospheric downturn predicted by our model occurs at $D \sim$ 10-20 cm, corresponding to projectile masses roughly equivalent to the mass of the atmospheric column the objects are traversing. This is well below the downturn observed in the fresh craters detected by CTX at $D = 4$-$5$ m (Daubar et al., 2013). The downturn in the CTX SFD is likely due to detection limits of CTX. The largest fresh crater, $D = 33.8$ m, observed, is consistent with our model and the prediction of the Hartmann isochron of one annual $D \sim 30$ m crater; however, fewer craters were observed than predicted by the isochron at smaller diameters. It is possible that $D < 30$ m craters are underrepresented in the observed CTX craters and continued observation, allowing for larger craters to form over a longer observation period, may potentially resolve this question.

**Acknowledgements**

J.-P. W. and A. P. were supported by a NASA Mars Data Analysis Program Grant Number NNX11AQ64G.

**References**

Aharonson, O. (2007) The modern impact cratering flux at the surface of Mars, *Lunar Planet. Sci. Conf.*, 38$^{th}$, #2288.

Aharonson, O., M. T. Zuber, and D. H. Rothman (2001) Statistics of Mars' topography from the Mars Orbiter Laser Altimeter' Slopes, correlations, and physical Models, *J. Geophys. Res.*, 106, 23,723–23,735.

Artmieva, N. A., and V. V. Shuvalov (2001) Motion of a fragmented meteoroid through the planetary atmosphere, *J. Geophys. Res.*, 106, 3297–3309.

Arvidson, R., G. Crozaz, R. J. Drozd, C. M. Hohenberg, C. J. Morgan (1975) Cosmic ray exposure ages of features and events at the Apollo landing sites, *Moon*, 13, 259–276.

Baldwin, B., and Y. Sheaffer (1971) Ablation and breakup of large meteoroids during atmospheric entry, *J. Geophys. Res.*, 76, 4653–4668.

Basilevsky, A. T., G. Neukum, S. C. Werner, A. Dumke, S. van Gasselt, T. Kneissl, W. Zuschneid, D. Rommel, L. Wendt, M. Chapman, J. W. Head, and R. Greeley (2009) Episodes of floods in MangalaValles, Mars, from the analysis of HRSC, MOC and THEMIS images, *Planet. Space Sci.*, 57, 917–943.

Bland, P. A. and T. B. Smith (2000) Meteorite accumulations on Mars, *Icarus*, 144, 21–26.

Bronshten, V. A. (1983) *Physics of meteoric phenomena*, Dordrecht, Reidel. 356 p.

Brown, P., R. E. Spalding, D. O. ReVelle, E. Tagliaferri, and S. P. Worden (2002) The flux of small near-Earth objects colliding with the Earth, *Nature*, 420, 294–296.

Burr, D. M., J. A. Grier, L. P. Keszthelyi, A. S. McEwen (2002) Repeated aqueous flooding from the Cerberus Fossae: Evidence for very recently extant, deep groundwater on Mars, *Icarus*, 159, 53–73.

Byrne S., C. M. Dundas, M. R. Kennedy, M. T. Mellon, A. S. McEwen, S. C. Cull, I. J. Daubar, D. E. Shean, K. D. Seelos, S. L. Murchie, B. A. Cantor, R. E. Arvidson, K. S. Edgett, A. Reufer, N. Thomas, T. N. Harrison, L. V. Posiolova, F. P. Seelos (2009) Distribution of Mid-Latitude Ground Ice on Mars from New Impact Craters, *Science*, 25, 1674–1676.

Ceplecha, Z., J. Borovička, W. G. Elford, D. O. Revelle, R. L. Hawkes, V. Porubčan, M. Šimek (1998) Meteor Phenomena and Bodies, *Space Sci. Rev.*, 84, 327–471.

Chyba, C. F., P. J. Thomas, and K. J. Zahnle (1993), The 1908 Tunguska explosion - Atmospheric disruption of a stony asteroid, *Nature*, 361, 40–44.

Daubar, I. J., and A. S. McEwen (2009) Depth to diameter ratios of recent primary impact craters on Mars, *Lunar Planet. Sci. Conf.*, 40$^{th}$, #2419.

Daubar, I. J., S. Byrne, A. S. McEwen, and M. Kennedy (2010) New Martian Impact Events: Effects on Atmospheric Breakup on Statistics, *1st Planetary Cratering Consortium*.

Daubar, I. J., A. S. McEwen, S. Byrne, C. M. Dundas, A. L. Keske, G. L. Amaya, M. Kennedy, and M. S. Robinson (2011) New craters on Mars and the Moon, *Lunar Planet. Sci. Conf.*, 42$^{nd}$, #2232.

Daubar I. J., A. S. McEwen, S. Byrne, M. R. Kennedy, and B. Ivanov (2013) The current martian cratering rate, *Icarus*, in press

Davis, P. M. (1993) Meteoroid impacts as seismic sources on Mars, *Icarus*, 105, 469–478.

Dundas, C. M., and S. Byrne (2010) Modeling Sublimation of Ice Exposed by Recent Impacts in the Martian Mid-Latitudes, *Icarus*, 206, 716–728.

Dundas C. M., L. P. Keszthelyi, V. J. Bray, and A. S. McEwen (2010) Role of material properties in the cratering record of young platy-ridged lava on Mars, *Geophys. Res. Lett.*, 37, L12203, doi:10.1029/2010GL042869.

Flynn, G. J., and D. S. McKay (1990) An assessment of the meteoritic contibution to the martian soil, *J. Geophys. Res.*, 95, 14,497–14,509.

Hartmann, W. K. (1966), Martian cratering, *Icarus*, 5, 565-576.

Hartmann, W. K. (1999), Martian cratering VI: crater count isochrons and evidence for recent volcanism from Mars Global Surveyor, *Meteor, Planet. Sci.*, 34, 167-177.

Hartmann, W. K. (2005), Martian cratering 8: Isochron refinement and the chronology of Mars, *Icarus*, 174, 294–320.

Hartmann, W. K. (2007), Martian cratering 9: toward resolution of the controversy about small craters, *Icarus*, 186, 274–278.

Hartmann, W. K. and D. C. Berman (2000), Elysium Planitia lava flows: crater count chronology and geological implications, *J.Geophys. Res.*, 105, 15,011–15,025.

Hartmann,W. K., G. Neukum, (2001) Cratering chronology and evolution of Mars. In: Altwegg, K., Ehrenfreund, P., Geiss, J., Huebner, W.F. (Eds.), *Composition and Origin of Cometary Materials*, Kluwer Academic, The Netherlands, pp. 165–194.

Hartmann, W. K., C. Quantin, S. C. Werner, and O. Popova (2010) Do young martian ray craters have ages consistent with the crater count system?, *Icarus*, 208, 621–635.

Hiesinger H., C. H. van der Bogert, J. H. Pasckert, L. Funcke, L. Giacomini, L. R. Ostrach, and M. S. Robinson (2012) How old are young lunar craters?, *J. Geophys. Res.*, 117, E00H10, doi:10.1029/2011JE003935.

Holsapple, K. A. (1993), The scaling of impact processes in planetary science, *Ann. Rev. Earth Planet. Sci.*, 21, 333–373.

Holsapple , K. A. and K. R. Housen (2007) A crater and its ejecta: An interpretation of Deep Impact, *Icarus*, 345–356.

Ivanov, B. A. (2006) Earth/Moon impact rate comparison: Searching constraints for lunar secondary/primary cratering proportion, *Icarus*, 183, 504–507.

Ivanov, B., H. J. Melosh, A. S. McEwen, and the HiRISE team (2008) Small impact crater clusters in high resolution HiRISE images, *Lunar Planet. Sci. Conf.*, 39$^{th}$, #1221.

Ivanov, B. A., H. J. Melosh, A. S. McEwen, and the HiRISE team (2009) Small impact crater clusters in high resolution HiRISE images – II, *Lunar Planet. Sci. Conf.*,40$^{th}$, #1410.

Ivanov, B. A., H. J. Melosh, A. S. McEwen, and the HiRISE team (2010) Small impact crater clusters in high resolution HiRISE images – III, *Lunar Planet. Sci. Conf.*,41$^{th}$, #2020.

Kadish, S. J., J. W. Head, R. L. Parsons, D. R. Marchant, (2008) The Ascraeus Mons fan-shaped deposit: Volcano–ice interactions and the climatic implications of cold-based tropical mountain glaciation, *Icarus*, 197, 84–109.

Kennedy, M. R., and M. C. Malin (2009) 100 new impact crater sites found on Mars, *AGU Fall Meeting*, #P43D-1455

Kneissel, T., S. van Gasselt, and G. Neukum (2011) Map-projection-independent crater size-frequency determination in GIS environments–New software tool for ArcGIS, *Planet. Space Sci.*, 59, 1243–1254.

Kovach, R. L., and J. S. Watkins (1973) The velocity structure of the Lunar crust, *The Moon*, 7, 63–75.

Lanagan, P. D., A. S. McEwen, L. P. Keszthelyi,, T. Thordarson (2001), Rootless cones on Mars indicating the presence of shallow equatorial ground ice in recent times, *Geophys. Res. Lett.*, 28, 2365–2367.

Love, S. G., and D. E. Brownlee (1991) Heating and thermal transformation of micrometeoroids entering the Earth's atmosphere, *Icarus*, 89, 26–43.

Malin, M. C., and K. S. Edgett (2000a) Evidence for recent groundwater seepage and surface runoff on Mars, *Science*, 288, 2330–2335.

Malin, M. C., and K. S. Edgett (2000b) Sedimentary rocks of early Mars, *Science*, 290, 1927–1937.

Malin, M. C., and K. S. Edgett (2001) Mars Global Surveyor Mars Orbiter Camera: Interplanetary cruise through primary mission, *J. Geophys. Res.*, 106, 23,429–23,570.

Malin, M. C., G. E. Danielson, A. P. Ingersoll, H. Masursky, J. Veverka, M. A. Ravine, and T. A. Soulanille (1992) Mars orbiter camera, *J. Geophys. Res.*, 97, 7699–7718.

Malin, M. C., K. S. Edgett, L. Posiolova, S. McColley, and E. Noe Dobrea (2006) Present impact cratering rate and the contemporary gully activity on Mars: Results of the mars global surveyor extended mission, *Science*, 314, 1573–1557.

Malin, M. C., J. F. Bell III, B. A. Cantor, M. A. Caplinger, W. M. Calvin, R. T. Clancy, K. S. Edgett, L. Edwards, R. M. Haberle, P. B. James, S. W. Lee, M. A. Ravine, P. C. Thomas, and M. J. Wolff (2007) Context Camera Investigation on board the Mars Reconnaissance Orbiter, *J. Geophys. Res.*, 112, E05S04, doi:10.1029/2006JE002808.

Mangold, N. (2003) Geomorphic analysis of lobate debris aprons on Mars at Mars Orbiter Camera scale: Evidence for ice sublimation initiated by fractures, *J. Geophys. Res.*, 108, 10.1029/2002JE001885, pp. GDS 2-1.

Marquez, A., Fernandez, C., Anguita, F., Farelo, A., Anguita, J., de la Casa, M.-A. (2004) New evidence for a volcanically, tectonically, and climatically active Mars, *Icarus*, 172, 573–581.

Marchi , S., S. Mottola, G. Cremonese, M. Massironi, and E. Martellato (2009) A new chronology for the Moon and Mercury, *Astron. J.*, 137, 4936–4948.

McEwen, A. S., and E. B. Bierhaus (2006) The importance of secondary cratering to age constraints on planetary surfaces, *Annu. Rev. Earth Planet. Sci.*, 34, 535–567.

McEwen, A. S., B. S. Preblich, E. P. Turtle, N. A. Artemieva, M. P. Golombek, M. Hurst, R. L. Kirk, D. M. Burr, and P. R. Christensen (2005), The rayed crater Zunil and interpretations of small impact craters on Mars, *Icarus*, 176, 351–381, doi:10.1016/j.icarus.2005.02.009.

McEwen, A. S., E. M. Eliason, J. W. Bergstrom, N. T. Bridges, C. J. Hansen, W. A. Delamere, J. A. Grant, V. C. Gulick, K. E. Herkenhoff, L. Keszthelyi, R. L. Kirk, M. T. Mellon, S. W. Squyres, N. Thomas, C. M. Weitz (2007a) Mars Reconnaissance Orbiter's High Resolution

Imaging Science Experiment (HiRISE), *J. Geophys. Res.*, 112, E05S02, doi:10.1029/2005JE002605.

McEwen, A. S., J. A. Grant, L. L. Tornabene, S. Byrne, and K. E. Herkenhoff (2007b) HiRISE observations of small impact craters on Mars, *Lunar Planet. Sci. Conf.*, 38$^{th}$, #2009.

McEwen, A. S., L. L. Tornabene, and the HiRISE Team (2007c) Modern Mars: HiRISE observations of small, recent impact craters on Mars, *7$^{th}$ Int. Conf. Mars*, LPI Contribution No. 1353, #3086.

Melosh, H. J. (1989), *Impact Cratering: A Geologic Process*, 245 pp., Oxford Univ. Press, New York.

Michael, G. G., and G. Neukum (2009) Planetary surface dating from crater size-frequency distribution measurements: Partial resurfacing events and statistical age uncertainty, *Earth Planet. Sci. Lett.*, 294, 223–229.

Neukum, G. and D. Wise, (1976) Mars: a standard crater curve and possible new time scale, *Science*, 194, 1381−1387.

Neukum, G., and B. A. Ivanov (1994), Crater Size Distributions and Impact Probabilities on Earth from Lunar, Terrestrial-planet, and Asteroid Cratering Data, in *Hazards Due to Comets and Asteroids*, edited by T. Gehrels, M. S. Matthews, and A. M. Schumann, pp. 359–416.

Neukum, G., B. A. Ivanov, and W. K. Hartmann (2001) Cratering records in the inner solar system in relation to the Lunar reference system, *Space Sci. Rev.*, 96, 55−86.

Podolak, M., J. B. Pollack, and R. T. Reynolds (1988) Interactions of planetesimals with protoplanetary atmospheres, *Icarus*, 73, 163–179.

Popova, O., I. Nemtchinov, and W. K. Hartmann (2003) Bolides in the present and past martian atmosphere and effects on cratering processes, *Meteorit. Planet. Sci.*, 38, 905–925.

Popova, O., W. K. Hartmann, I. Nemtchinov, D. C. Richardson, and D. C. Berman (2007), Crater clusters on Mars: shedding light on martian ejecta launch conditions, *Icarus*, 190, 50–73.

Popova, O., J. Borovička, W. K. Hartmann, P. Spurný, E. Gnos, I. Nemtchinov, and J. M. Trigo-Rodríguez (2011) Very low strengths of interplanetary meteoroids and small asteroids, *Meteorit. Planet. Sci.*, 46, 1525–1550.

Quantin, C., N. Mangold, W. K. Hartmann, and P. Allemand (2007), Possible long-term decline in impact rates 1. Martian geological data, *Icarus*, 186, 1–10.

Reiss, D., S. van Gasselt, G. Neukum, and R. Jaumann (2004) Absolute dune ages and implications for the time of formation of gullies in Nirgal Vallis, Mars, *J. Geophys. Res*. 109, E06007, doi:10.1029/2004JE002251.

Richardson, J. E., H. J. Melosh, C. M. Lisse, and B. Carcich (2007) A ballistic analysis of the Deep Impact ejecta plume: Determining comet Tempel 1's gravity, mass, and density, *Icarus*, 190, 357–390.

Schmidt, R. M., and K. R. Housen (1987), Some recent advances in the scaling of impact and explosion cratering , *Int. J. Impact Eng.*, 5, 543–560.

Schon, S. C., J. W. Head, and C. I. Fassett (2009) Unique chronostratigraphic marker in depositional fan stratigraphy on Mars: Evidence for ca. 1.25 Ma gully activity and surficial meltwater origin, *Geology*, 37, 207–210.

Scott, D. H., J. M. Dohm, and J. R. Zimbelman (1998) Geologic map of Pavonis Mons volcano, Mars, US Geol. Surv. Misc. Invest. Ser. Map I-2561.

Shean, D. E., J. W. Head, M. Kreslavsky, G. Neukum and HRSC Co-I Team (2006) When were glaciers present in Tharsis? Constraining age estimates for the Tharis Montes fan-shaped deposites, *Lunar Planet. Sci. Conf.*, 37$^{th}$, #2092.

Smith, D. E., M. T. Zuber, S. C. Solomon, R. J. Phillips, J. W. Head, J. B. Garvin, W. B. Banerdt, D. O. Muhleman, G. H. Pettengill, G. A. Neumann, F. G. Lemoine, J. B. Abshire, O. Aharonson, C. D. Brown, S. A. Hauck, A. B. Ivanov, P. J. McGovern, H. J. Zwally, T. C. Duxbury (1999) The global tpography of Mars and implications for surface evolution, *Science*, 284, 1495–1503.

Svetsov, V. V., I. V. Nemtchinov, and A. V. Teterev (1995) Disintegration of large meteoroids in Earth's atmosphere: Theoretical models, *Icarus*, 116, 131–153.

Vasavada, A. R., J. L. Bandfield, B. T. Greenhagen, P. O. Hayne, M. A. Siegler, J.-P. Williams, and D. A. Paige (2012) Lunar equatorial surface temperatures and regolith properties from the Diviner Lunar Radiometer Experiment, *J. Geophys. Res.*, 117, E00H18, doi:10.1029/2011JE003987.

Watkins, J. S. and R. L. Kovach (1973) Seismic investigation of the lunar regolith, *Proc. Fourth Lunar Sci. Conf.*, 3, 2561–2574.

Weibull, W. A. (1951) A statistical distribution function of wide applicability, *J. Appl. Mech.*, 10, 140–147.

Wilhelms, D. E. (1987) The geologic history of the Moon, *U.S. Geol. Surv. Prof. Pap.*, 1348, 302 pp.

Williams, J.-P., O. Aharonson, and A. V. Pathare (2010) The production of small primary craters on Mars, *Lunar Planet. Sci. Conf.*, 41$^{st}$, #2574.

**Tables**

**Table 1**
Distribution of material types based on fireball network observations (Ceplecha et al., 1998).

| Group | % obs. | Density, $\rho_m$ (kg m$^{-3}$) | Ablation Coef., $\sigma$ |
| --- | --- | --- | --- |
| Irons | 3 | 7800 | $7.0 \times 10^{-8}$ |
| Ordinary Chondrites | 29 | 3700 | $1.4 \times 10^{-8}$ |
| Carbonaceous Chondrites | 33 | 2000 | $4.2 \times 10^{-8}$ |
| Cometary Material | 26 | 750 | $10.0 \times 10^{-8}$ |
| Soft Cometary Material | 9 | 270 | $21.0 \times 10^{-8}$ |

**Table 2**
Sensitivity analysis

| Model (1 yr) | Best-fit[1] isochron from Hartmann (2005) | |
| --- | --- | --- |
|  | $D = 4 – 20$ m | $D = 20 – 300$ m |
| Nominal | $0.734 \pm 0.002$ | $1.030 \pm 0.017$ |
| Highlands ($\rho_o = 0.0135$ kg m$^{-3}$) | $0.920 \pm 0.002$ | $1.140 \pm 0.018$ |
| Lowlands ($\rho_o = 0.0222$ kg m$^{-3}$) | $0.687 \pm 0.002$ | $0.993 \pm 0.017$ |
| $\sigma \times \frac{1}{2}$ | $0.946 \pm 0.002$ | $1.170 \pm 0.018$ |
| $\sigma \times 2$ | $0.511 \pm 0.002$ | $0.843 \pm 0.016$ |
| Ordinary Chondrite only | $1.75 \pm 0.003$ | $2.13 \pm 0.025$ |
| Carbonaceous Chondrite only | $0.46 \pm 0.002$ | $0.872 \pm 0.017$ |
| Cometary Marterial only | $0.047 \pm 0.001$ | $0.15 \pm 0.007$ |
| Lunar Regolith | $0.903 \pm 0.002$ | $0.959 \pm 0.016$ |
| Soft Rock | $0.327 \pm 0.001$ | $0.561 \pm 0.013$ |
| Hard Rock | $0.193 \pm 0.001$ | $0.295 \pm 0.009$ |

[1] Isochrons fit to cumulative SFD using Craterstats2 program (Michael and Neukum, 2009)

**Figures**

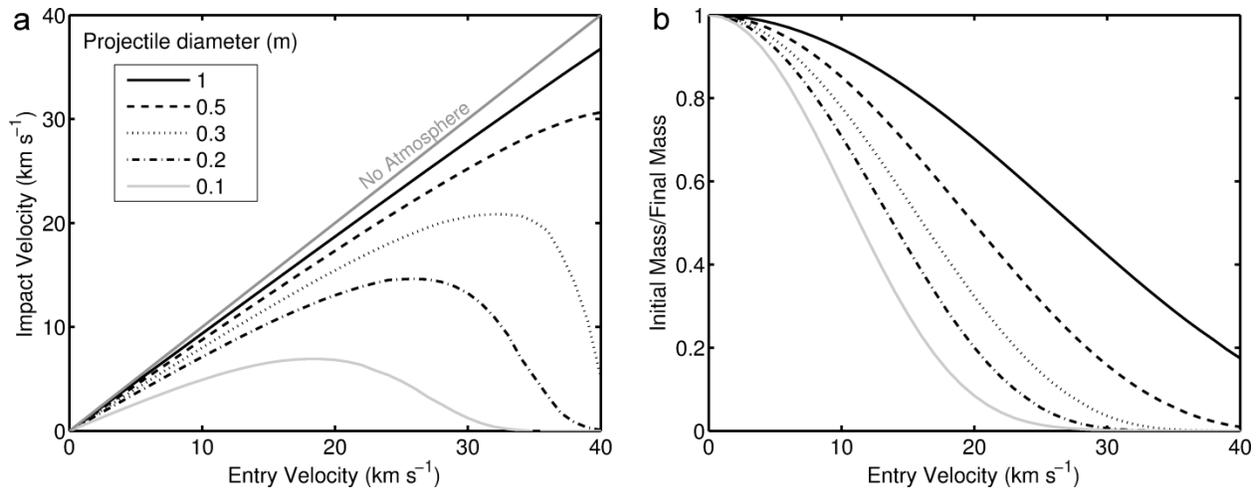

**Figure 1.** (a) Initial velocity versus final velocity and (b) the ratio of initial and final projectile mass versus initial velocity for a range of projectile diameters (10 cm – 1 m) where objects have properties of ordinary chondrites (Table 1). Smaller, faster objects are more effectively decelerated and ablated.

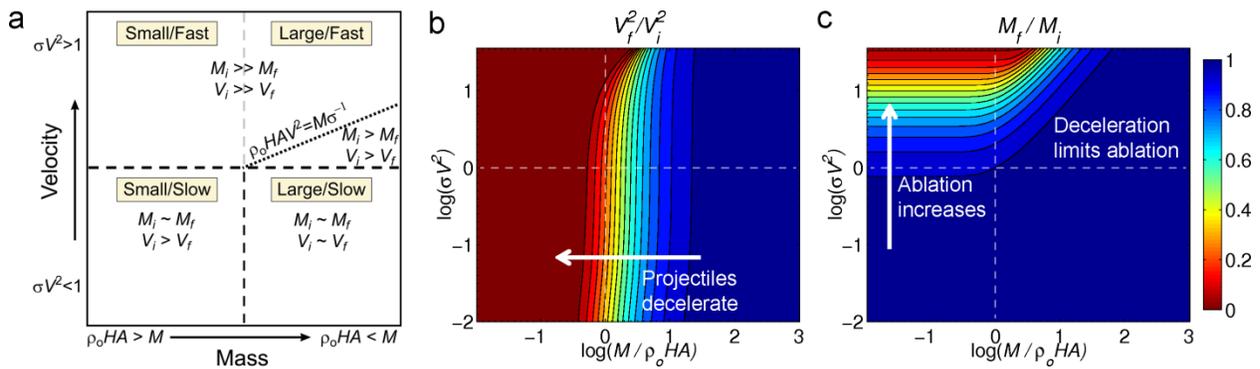

**Figure 2.** (a) General classes of projectiles based on initial mass and velocity. *Large, slow projectiles*: remain relatively unchanged as they are not energetic enough to ablate and too massive for the atmosphere to decelerate. *Small, slow projectiles*: the mass of the atmospheric column becomes comparable to the mass of the object and deceleration occurs. *Fast projectiles*: deceleration and ablation are significant as the energy of traversing a column of atmosphere at a given velocity exceeds that required to ablate the entirety of its mass. The triangular wedge defines intermediate objects that would significantly ablate if deceleration does not occur, however deceleration limits the ablation. Model results of the ratios of initial and final (b)velocities and (c) masses are shown for projectile masses non-dimensionalized by the mass of the atmospheric column encountered on the x-axis and the initial projectile velocity squared by the energy per unit mass, $\sigma^{-1}$, on the y-axis. For $\sigma = 1 \times 10^{-8}$ kg J$^{-1}$, slow projectiles have $v_i < 10$ km s$^{-1}$.

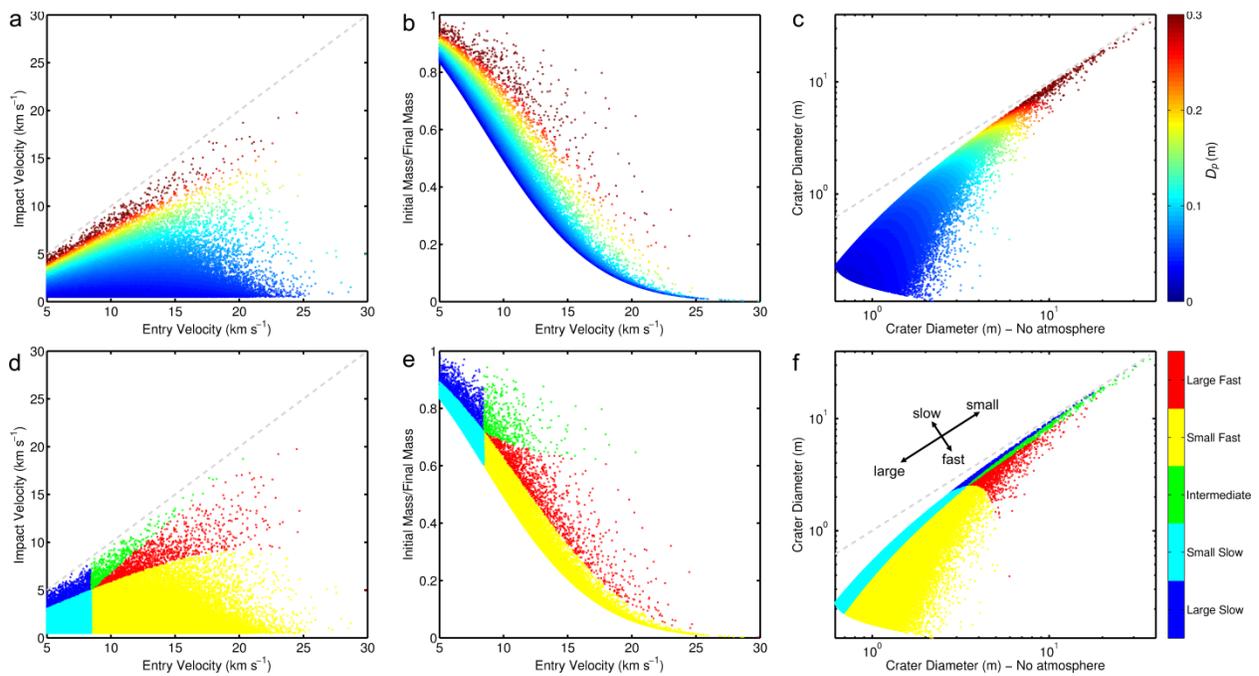

**Figure 3.** Scatter plots of results from the Monte Carlo simulation with a bivariate distribution of velocities and projectile diameters for $2 \times 10^5$ events with initial entry angles 45° and ordinary chondritic compositions. (a) Impact velocity versus initial entry velocity, (b) initial and final mass ratio versus entry velocity, and (c) final crater diameter versus crater diameter without an atmosphere, i.e. crater diameter resulting from initial mass and velocity. The color scale of (a-c) shows the initial projectile diameter. (d-f) are the same as (a-c) but events are tagged with 5

colors representing the general class of projectiles as defined in figure 2 (Slow/Fast, Small/Large). This trend is show with arrows in (f). Deviations from the grey dashed lines in (a),(c),(d), and (f), and from the horizontal in (b) and (e), shows the magnitude of ablation and/or deceleration experienced by the objects. Crater diameters are determined assuming parameters of dry soil with effective strength, $\bar{Y} = 65$ kPa and $\rho_t = 2000$ kg m$^{-3}$ (Holsapple, 1993).

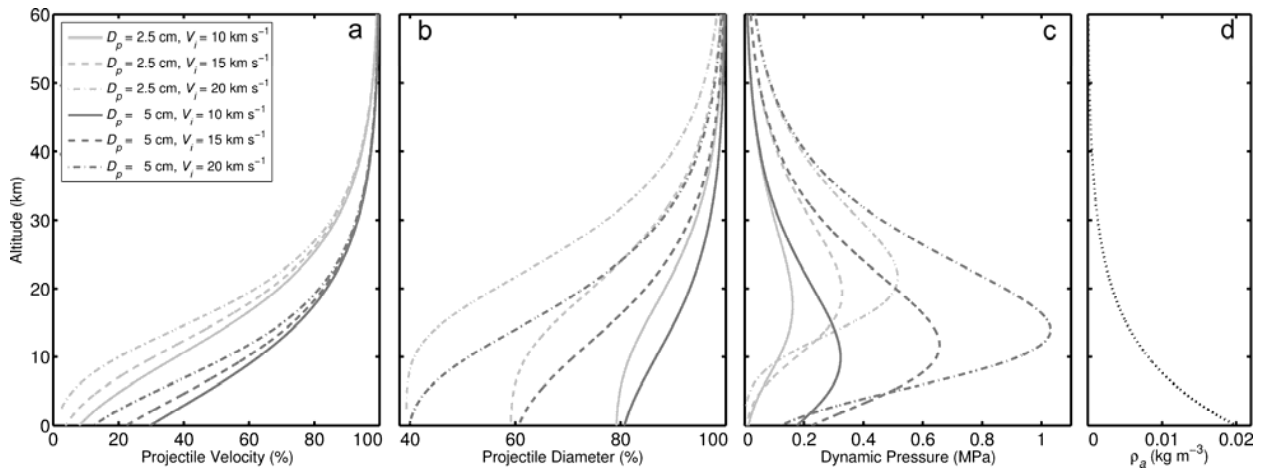

**Figure 4.** Profiles of (a) the percent of initial velocity, (b) the percent of initial diameter, and (c) the dynamic pressure for six ordinary chondrites in the martian atmosphere as a function of altitude with entry angles 45°. (d) Assumed atmospheric density with altitude. Note that while the faster projectiles decelerate to a greater relative extent, they experience larger dynamic pressure upon entry. Ablation is more sensitive to velocity, however, deceleration is more sensitive to projectile size, and therefore smaller objects tend to experience lower dynamic pressures.

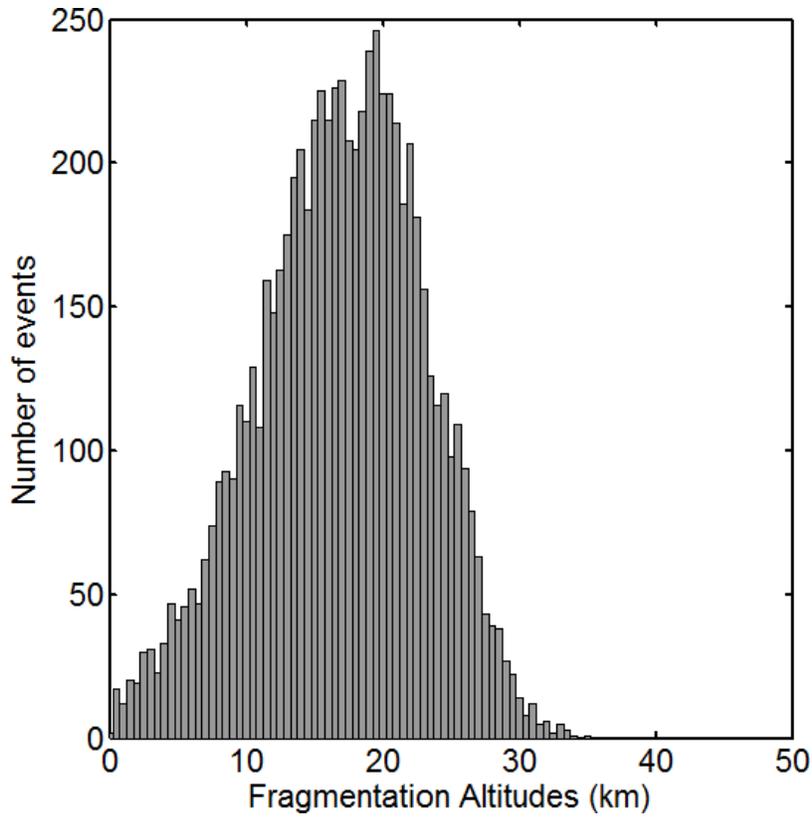

810

811 **Figure 5.** Histogram of fragmentation altitudes using $\sigma_m$ = 0.65 MPa. Most fragmentation occurs

812 ~2-3 scale heights above the surface with few fragmentation events occurring at lower altitudes.

813

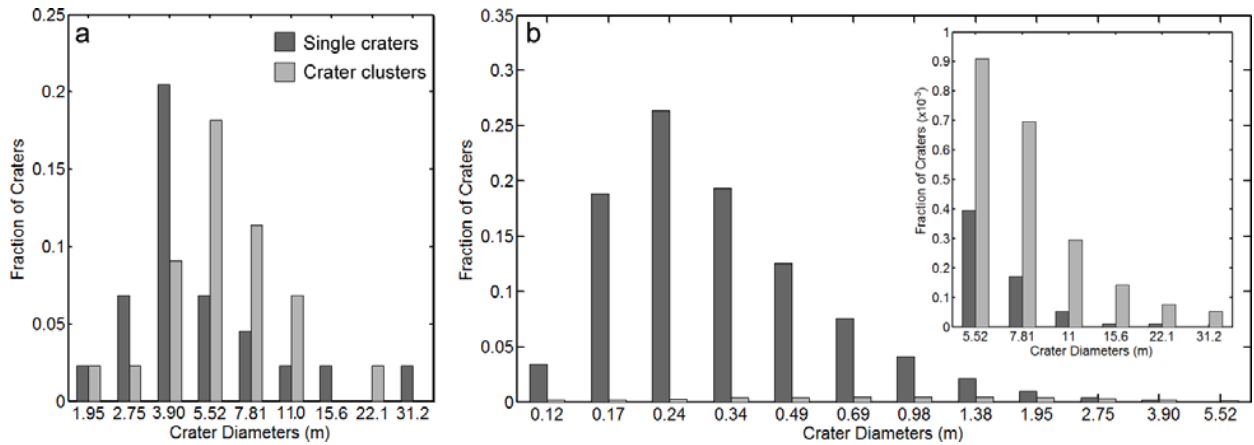

814

**Figure 6.** Histograms of crater diameters for (a) the 44 fresh craters reported by Daubar et al. (2013) and (b) the model using $\sigma_m = 0.65$ MPa. Diameters of crater clusters are effective diameters (see text). The inset is the model craters for $D > 5$ m.

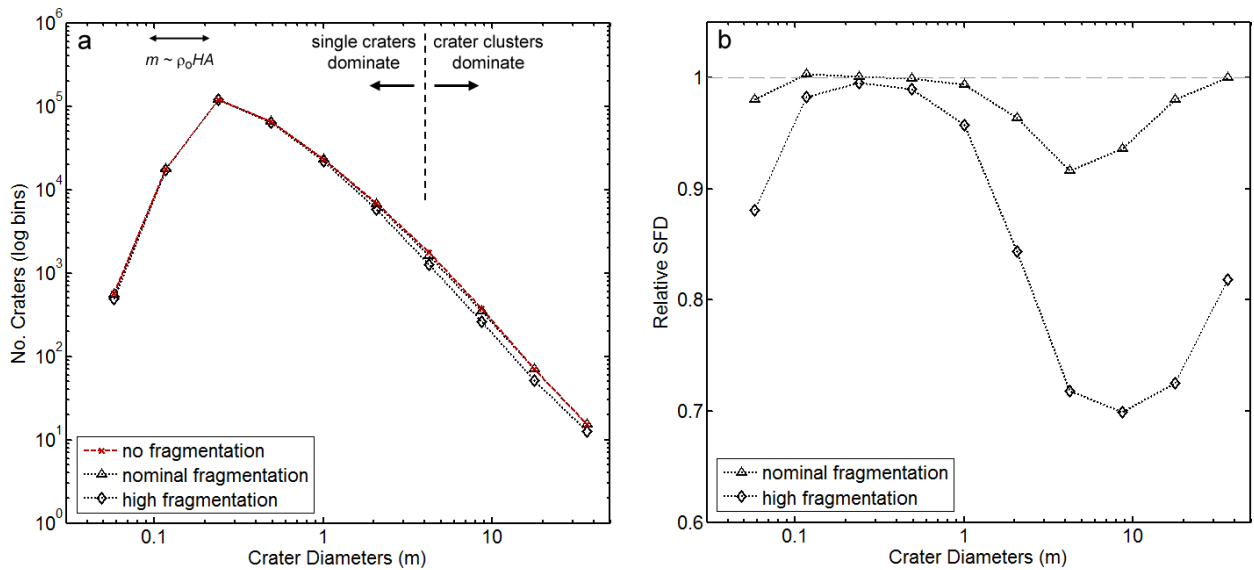

**Figure 7.** (a) The modeled crater SFDs where no fragmentation occurs (i.e. $\sigma_m \gg \rho_a v^2$ for all meteoroids) and two different fragmentation characteristics: one favoring fewer fragments with few large fragments and many small fragments (nominal model), and one with a uniform random distribution of fragment numbers and sizes to increase the influence of fragmentation on the crater SFD. (b) The ratio of crater SFDs with and without fragmentation. The influence of fragmentation on the SFD is greatest at $D \sim 2 - 20$ m.

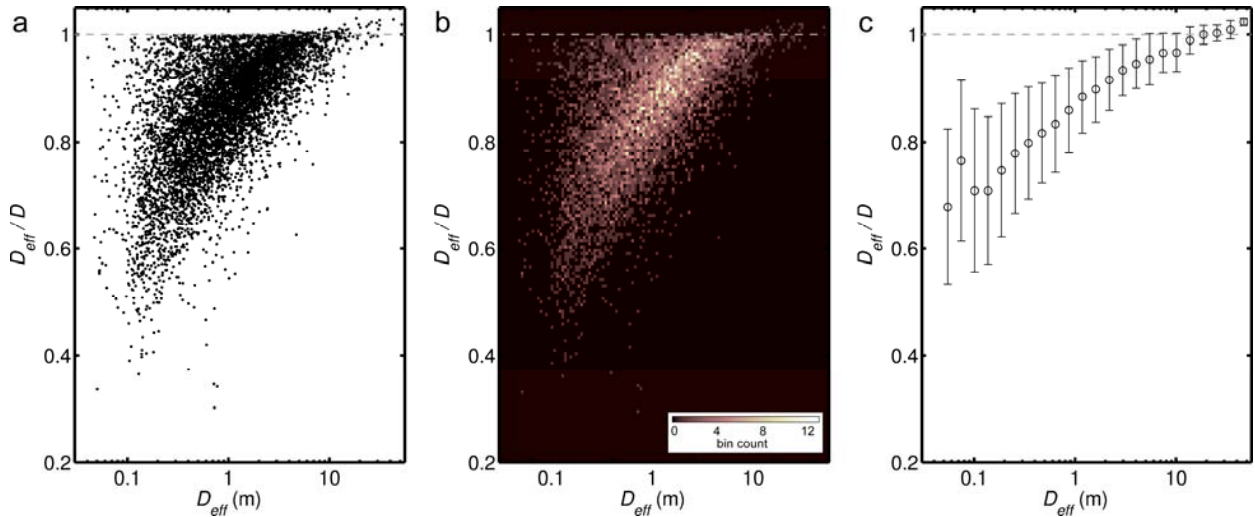

828

829  **Figure 8.** Model $D_{eff}/D$ for: (a) individual crater clusters (b) a 2D histogram and (c) the mean in

830  log $D_{eff}$ bins. Error bars represent the standard deviation in each bin. $D_{eff}$ is increasingly

831  unreliable at smaller sizes below ~10 m.

832

833

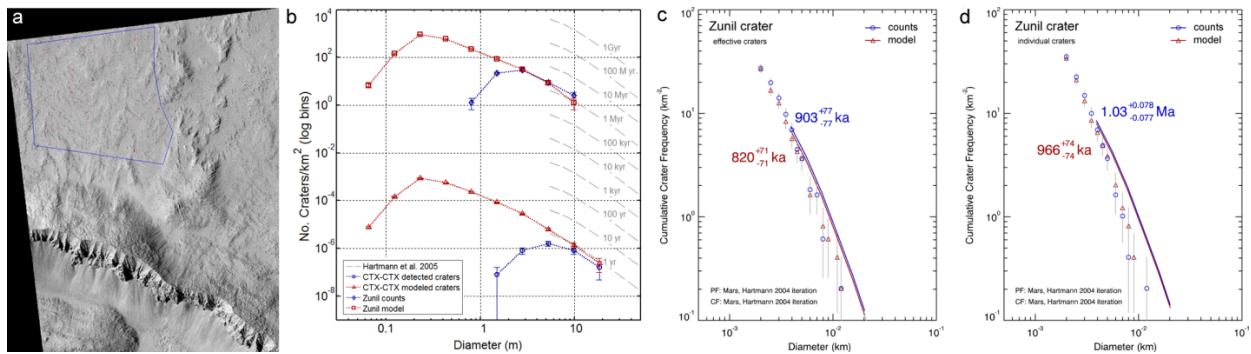

834

835  **Figure 9.** Crater counts conducted on a ~5 km² area north of the Zunil crater rim using HiRISE

836  image PSP_001764_1880 which was calibrated and map-projected using ISIS (Integrated

837  Software for Imagers and Spectrometers) and imported into Arcmap, with CraterTools (Kneissl

838  et al., 2011) used to identify and measure crater diameters. (a) Location of Zunil crater count

839  area in HiRISE image PSP_001764_1880. (b) Log-differential plot of $D_{eff}$ for crater counts for

840  Zunil and the 44 new impact sites constrained by CTX (Daubar et al., 2013) (blue) and model

841    results (red) for the corresponding surface area and time. Cumulative crater frequency of (c) $D_{eff}$

842    and (d) for individual craters for crater counts (blue) and model results (red) using the

843    Craterstats2 software tool (Michael and Neukum, 2009).

844

845

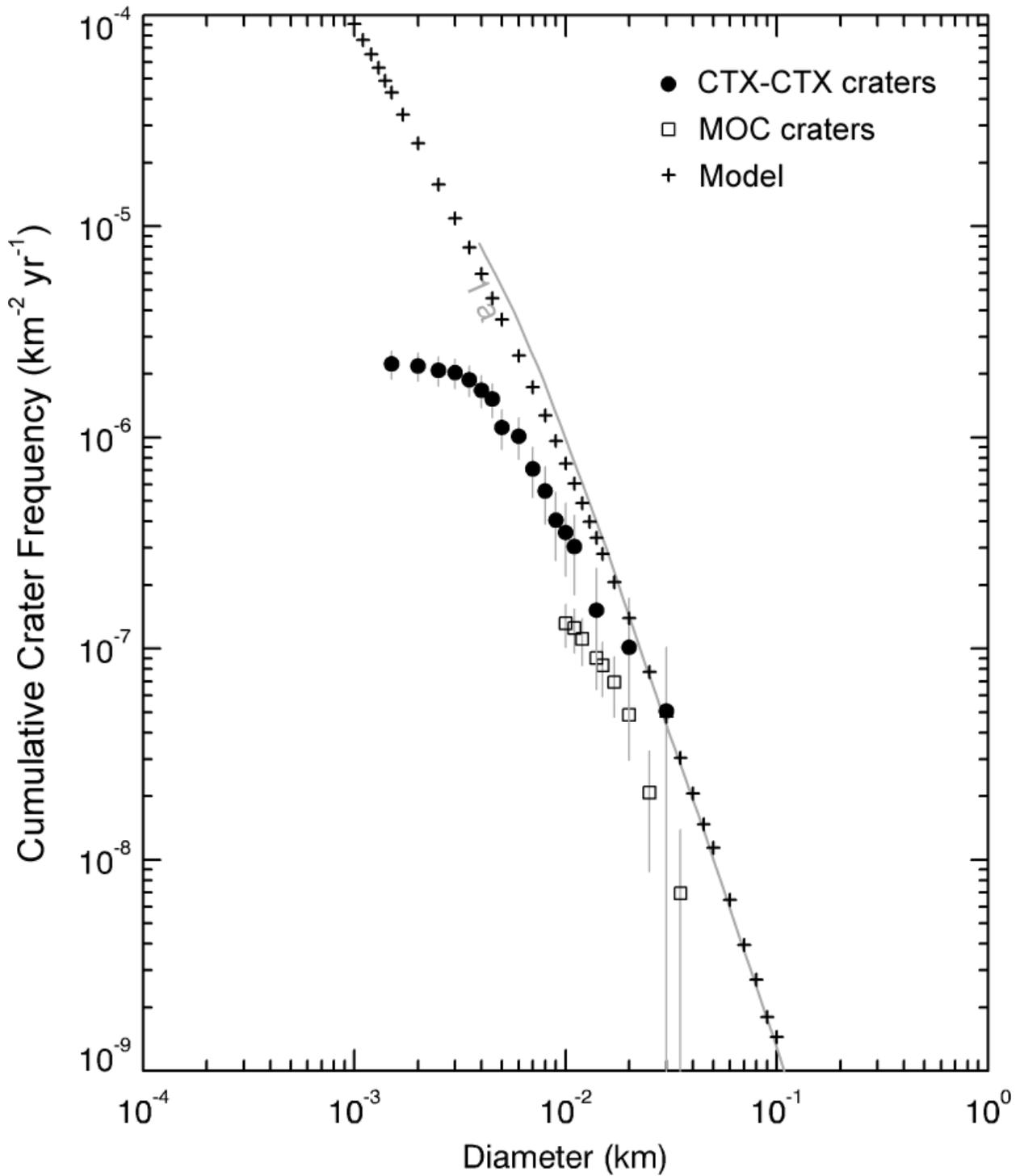

846

847 **Figure 10.** The cumulative SFD for the fresh craters from CTX-CTX detections (Daubar et al.,

848 2013), MOC (Malin et al., 2006), and model results, scaled to the same time/area, and the annual

849 Hartmann (2005) isochron (gray line).

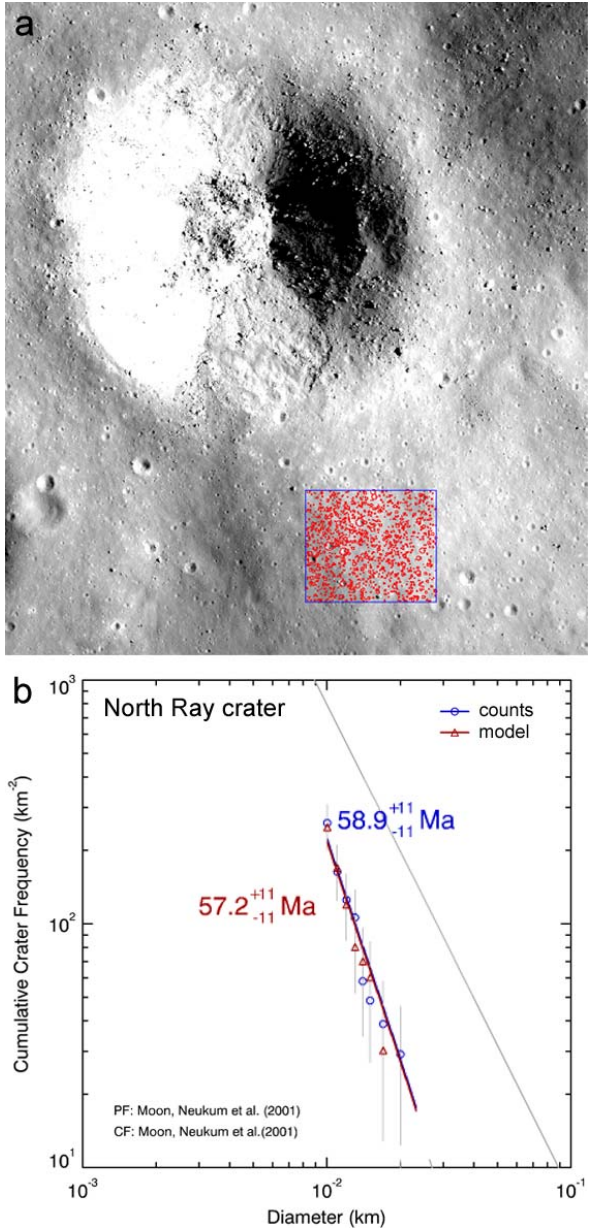

**Figure 11.** (a) Location of North Ray crater count area (350 m × 300 m blue box) in LRO NAC image M129187331 (NASA/GSFC/Arizona State Univ.). (b) The cumulative crater frequency of the crater counts (blue) and the model (red) using the Craterstats2 software tool (Michael and Neukum, 2009). The gray line indicates crater saturation.

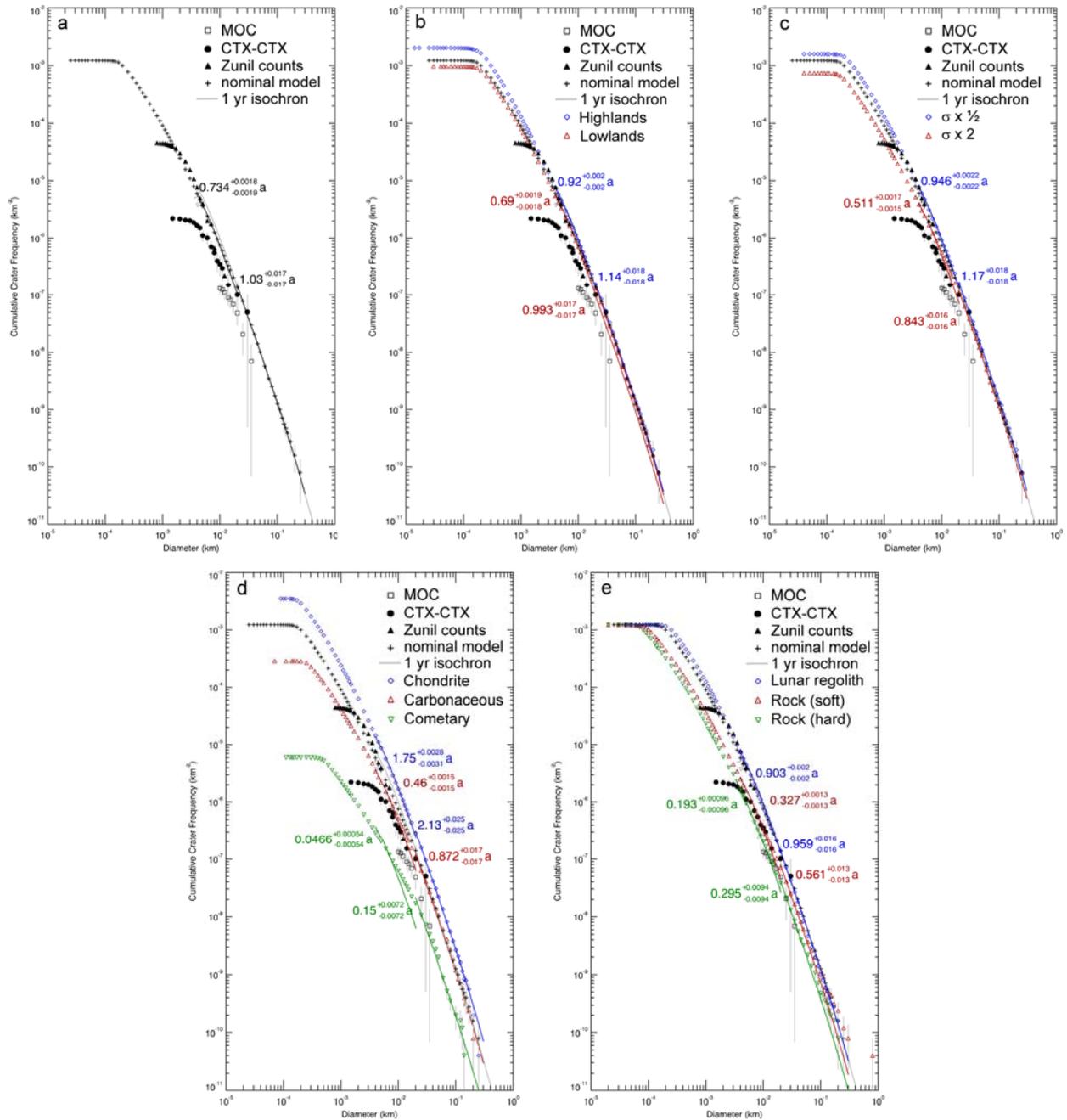

**Figure 12.** Sensitivity of cumulative SFD to model parameters. (a) Nominal model, fresh craters identified with MOC (Malin et al., 2006), and CTX (Daubar et al., 2013), and counts conducted on Zunil ejecta scaled to an annual isochron of Hartmann (2005). Estimated ages are for $D = 4 - 20$ m and $D = 20 - 300$ m. Nominal model assumes: $\rho_o = 0.02$ kg m$^{-3}$ consistent with the elevation of Zunil crater (-2.8 km), the distribution of meteoroid types with corresponding

ablation coefficients as given by Ceplecha et al. (1998), and target material properties consistent with dry desert alluvium, $K_1$= 0.24, $\mu$ = 0.41 and $\bar{Y}$ = 65 kPa. (b) Results for atmospheric surface densities for the Highlands and Lowlands. (c) Results scaling the ablation coefficients, $\sigma$, by 0.5 and 2. (d) Results assuming all projectiles made of ordinary chondrites, carbonaceous chondrites, or cometary material as defined in Table 1. (e) Results assuming target material properties of lunar regolith, soft rock, and hard rock.